"Thornhill, de Broglie and the kinetic theory of electromagnetic radiation"

J. P. Dunning-Davies and J. Dunning-Davies,
Department of Physics,
University of Hull,
Hull HU10 7UN,
England.

email: j.dunning-davies@hull.ac.uk

**Abstract.**

In 1983, Thornhill showed that Planck's energy distribution for a black-body radiation field could be derived for a gas-like æther with Maxwellian statistics. Further it was shown that the frequency of electromagnetic waves correlates with the energy per unit mass of the particles, not with their energy, thus differing from Planck's quantum hypothesis. He pointed out that de Broglie, in a paper of 1922, was on the way to achieving this result but did not pursue the matter to its logical conclusion. Here a translation of de Broglie's paper is presented to draw attention to this point. Some brief additional comments are included also.

## Introduction.

In 1983, Thornhill[1] showed that Planck's energy distribution for a black-body radiation field could be derived for a gas-like æther with Maxwellian statistics. Further it was shown that the frequency of electromagnetic waves correlates with the energy per unit mass of the particles, not with their energy, thus differing from Planck's quantum hypothesis. He pointed out that de Broglie, in a paper of 1922[2], was on the way to achieving this result but did not pursue the matter to its logical conclusion. A translation of de Broglie's paper follows, together with some additional comments.

## Translation of de Broglie's paper:

<div style="text-align:center">

"On the interference and theory of light quanta"
[Comptes Rendus **175** (1922)811]

</div>

Recent advances in physics in the field of emission and absorption of radiation draw more and more attention to the theory of light quanta according to which energy from all forms of radiation (Hertzian, luminous, X or γ) would be concentrated in small individual elements, equal to $h\nu$, constituting, after a fashion, 'atoms of light' of frequency $\nu$. In certain circumstances, these atoms of light would combine together into molecules. The explanation behind the theory of the phenomenon of light quanta, interpreted up to this point by the hypothesis of waves such as those from interference, diffusion, dispersion, and so on, seems quite tedious but, if managed well, it should be possible to compromise between the old theory and the new by introducing the notion of periodicity. At the time of this synthesis, Maxwell's equations would have appeared without doubt to be a continuous approximation (possibly a valid argument in some cases, but not all) of the discontinuous structure of radiant energy, just as the continuum equations of hydrodynamics satisfactorily represent the movement of fluids on the macroscopic scale where the atomic structure is not being considered.

At this point, perhaps we should consider the idea of developing a theory of interference in harmony with the existence of light quanta.

It is known that the energy fluctuations in black body radiation held in thermal equilibrium in a volume $V$ are governed by

$$\overline{\varepsilon^2} = kT^2 \frac{dE}{dT},$$

where $T$ is the temperature of the enclosure, $k$ is Boltzmann's constant, $\varepsilon$ is the deviation from the average value $E$ of the instantaneous value of the energy at frequency $\nu$ in the spectral interval $d\nu$ in the volume $V$.

If, firstly, one supposes that black body radiation is governed by the Raleigh-Jeans law, $E = \frac{8\pi k}{c^3} \nu^2 TVd\nu$, one finds

$$\overline{\varepsilon^2} = \frac{c^3}{8\pi v^2 dv} \times \frac{E^2}{V},$$

and this result, as one would expect, coincides with the result, derived via the theory of electromagnetism, for the interference of black body radiation.

If we adopt Wien's law as the distribution law

$$E = \frac{8\pi h}{c^3} v^3 e^{-hv/kT} V dv,$$

which corresponds to the hypothesis of radiation entirely divided into quanta $hv$, one finds $\overline{\varepsilon^2} = hvE$, readily indicating the direct reasons for the fluctuations of light quanta.

In fact, in the realistic case of Planck's law

$$E = \frac{8\pi h}{c^3} v^3 \frac{1}{e^{hv/kT} - 1} V dv,$$

one finds, as Einstein showed at the Congress of Brussells in 1911,

$$\overline{\varepsilon^2} = hvE + \frac{c^3}{8\pi v^2 dv} \frac{E^2}{V};$$

$\overline{\varepsilon^2}$ is, therefore, the sum of two terms – (i) if the radiation was purely wave like and (ii) if the radiation was entirely split up into quanta $hv$.

From the point of view of the theory of light quanta, it seems logical to write Planck's formula in the form:

$$E = \frac{8\pi h}{c^3} v^3 e^{-hv/kT} V dv + \frac{8\pi h}{c^3} v^3 e^{-2hv/kT} V dv + \frac{8\pi h}{c^3} v^3 e^{-3hv/kT} V dv + ...$$
$$= \sum_{n=1}^{\infty} \frac{8\pi h}{c^3} v^3 e^{-nhv/kT} V dv = E_1 + E_2 + E_3 + ....$$

The first term $E_1$ corresponds to the energy divided into quanta $hv$, the second $E_2$ corresponds to the energy divided into quanta $2hv$, and so on. Then, the formula for fluctuations is given by

$$\overline{\varepsilon^2} = hvE_1 + 2hvE_2 + 3hvE_3 + ... = \sum_{n=1}^{\infty} nhvE_n,$$

and this formula also agrees well with a 'gas of light' composed of molecules and atoms. Naturally, this new idea is really identical at heart with that of Einstein in that it is easy to verify that

$$\sum_{n=2}^{\infty}(n-1)h\nu E_n = \frac{c^3}{8\pi\nu^2 d\nu}\frac{E^2}{V}.$$

If one were to study these formulæ in depth, one would see they have the following significance: from the point of view of light quanta, the phenomena of interference are consistent and appear bound to the existence of the conglomeration of atoms of light moving coherently, rather than independently. Consequently, it is natural to assume that, if one day the theory of light quanta manages to explain the interference, it should be possible to explain this conglomeration of quanta.

Reference.

[1] See Lorentz, "*Statistical theories in thermodynamics*", (Conference in the Collège de France), edited by M. Dunoyer, Hermann, p.71.

**Some comments.**

If anything, it is the equations in the above translation of de Broglie's work which are of significance. In his 1983 article, Thornhill[1] draws attention to the implications of the above article. He notes that de Broglie considered the possibility of an infinite variety of light quanta, each having an energy which was an integral multiple of Planck's energy quantum *hv*. Further, he noted that Planck's distribution could be expanded as an infinite series in $\nu^3 e^{-nh\nu/kT}$, each term having the same form as Wien's distribution. de Broglie drew attention to the expression Einstein had derived earlier for the mean square energy fluctuations per unit volume, expanded this result as an infinite series of terms, and showed that it corresponded, term by term, with the fluctuations found individually for the Wien-like terms in the expansion of Planck's distribution. Hence, the terms of each series might be thought to correspond to energy quanta *nhv*, which indicated the possibility of obtaining both Planck's distribution and Einstein's fluctuation formula on the basis of a particulate theory of electromagnetic radiation – always assuming a suitably weighted mixture could be found for these different corpuscles of energies *nhv*. Although such a corpuscular theory had been suggested earlier by Wolfke[2], it seems de Broglie didn't pursue this line of investigation any further. Apparently, Bothe[3], who quoted Wolfke's work but must have been unaware of de Broglie's contribution since he did not refer to it, pursued the matter much further but not to its logical conclusion and it was left to Thornhill to complete the task[1]. It seems that this article by Thornhill is deserving of more serious open-minded perusal by the scientific community in that it makes some interesting, if controversial, predictions. Possibly the most important of these predictions is that the speed of light varies with the square root of the background temperature. If this is so, the need for a theory of inflation disappears immediately and other consequences would follow undoubtedly.

This discussion generally draws attention to the fact that there is much to be learnt, even today, from the original writings and thoughts of the pioneers of our modern ideas in theoretical physics. Much of this is written in French and German and is, therefore, unfortunately inaccessible to many today. Hopefully, this short contribution will alert people to this wealth of hidden scientific information and result in more of it being made available, through translation, to the wider scientific community.

**References.**